\documentclass{article}

\usepackage{graphicx}
\usepackage{fullpage,amsmath,amssymb,latexsym}
\usepackage{multirow}
\usepackage{natbib}
\begin{document}
\title{Interior Models of Uranus and Neptune}
\author{Ravit Helled$^1$, John D. Anderson$^2$, Morris Podolak $^3$, and Gerald Schubert$^1$\\
\small{$^1$Department of Earth and Space Sciences and Institute of Geophysics and Planetary Physics,}\\
\small{University of California, Los Angeles, CA 90095Ð1567, USA}\\
\small{$^2$Jet Propulsion Laboratory \footnote{Retiree},} 
\small{California Institute of Technology, Pasadena, CA 91109}\\
\small{$^3$ Tel Aviv University, Dept. of Geophysics and Planetary Science,} 
\small{Tel Aviv, Israel}\\
\small{corresponding author {\bf R. Helled: rhelled@ucla.edu}}\\
}
\date{}
\maketitle 

\begin{abstract}
'Empirical' models (pressure vs. density) of Uranus and Neptune interiors constrained by the gravitational
coefficients J$_2$, J$_4$, the planetary radii and masses, and Voyager solid-body rotation periods are presented.
The empirical pressure-density profiles are then interpreted in terms of physical equations of state of hydrogen, helium, ice (H$_2$O), and rock (SiO$_2$) to test the physical plausibility of the models. The compositions of Uranus and Neptune are found to be similar with somewhat different distributions of the high-Z material. The big difference between the two planets is that Neptune requires a non-solar envelope while Uranus is best matched with a solar composition envelope. Our analysis suggests that the heavier elements in both Uranus' and Neptune's interior might increase gradually towards the planetary centers.  Indeed it is possible to fit the gravitational moments without sharp compositional transitions.
\end{abstract}


\section{Introduction}
Uranus and Neptune, the ice giants, are the most distant planets in the solar system and have masses of 14.536 and 17.147 Earth masses, respectively. Although their interior structures are poorly understood, some inferences are possible based on their masses, radii, and gravitational fields as measured by Voyager 2 (Smith et al. 1986; 1989). The data suggest that these ice giants contain hydrogen and helium atmospheres, but unlike the gas giant planets, their hydrogen/helium mass fractions are small. The composition below the atmospheric layer is unknown but it is commonly assumed to be a mixture of rocks and ices (Hubbard et al. 1991; Podolak et al. 1995).  \par

There is a fundamental difference between models of Jupiter 
and Saturn and models of Uranus and Neptune. The former have massive gas envelopes 
and relatively small cores. A 'core' is a region of heavy elements at the center of the planet distinct in composition from the overlying envelope and separated from it by a density discontinuity. In fact, Jupiter may have no core at all as suggested by recent interior models that imply core masses between 0 and 6 Earth masses (Saumon and Guillot 2004). Interior models have shown that even Saturn's core mass could be very small and possibly zero (Guillot 1999). \par

In the case of Jupiter and Saturn, the gravitational moments are not sensitive to the presence of a 'core' so core mass and composition are not well constrained (Podolak and Hubbard 1998; Saumon and Guillot 2004). For 
Uranus and Neptune, however, the heavier elements represent a much larger fraction of the planetary mass, and in addition, the inner region is better sampled by J$_2$, the gravitational quadrupole moment.  
Figure 1 presents the normalized integrands of the gravitational moments (contribution functions) for Jupiter and Neptune. The functions describe how the different layers in the planets' interiors contribute to the total mass and the gravitational moments (Zharkov and Trubitsyn 1974; Guillot and Gautier 2007). The core region in Jupiter could extend out to $\sim$ 15 \% of the planet's radius and is not well sampled by the gravitational moments. For Neptune on the other hand, the core region could extend out to 70\% of the planet's radius, a region that is well sampled by the gravitational harmonics. In addition to the larger core region in Neptune compared with Jupiter, the peaks of the contribution functions occur at greater depth in Neptune, a property that favors sensitivity of the harmonics to the inner structure of Neptune-like planets. The contribution functions shown in Figure 1 were calculated with a density profile represented by a 6$^{th}$ order polynomial and could change slightly if different interior density models are used. 
Though we show results only for Jupiter and Neptune in Figure 1, the plots similarly apply to Saturn and Uranus as well. \par

The available constraints on interior models of Uranus and Neptune are limited. The gravitational harmonics of these planets have been measured only up to fourth degree (J$_2$, J$_4$), and the planetary shapes and rotation periods are not well known (Helled et al. 2010). 
The thermal structure of these planets also presents difficulties. Hubbard (1968) showed 
that Jupiter's high thermal emission implies a thick convective envelope and a thermal gradient that is close to adiabatic. Saturn and Neptune also have large thermal fluxes, so 
an adiabatic thermal gradient might be a good approximation for these planets as well. However, Uranus' thermal emission is close to zero (Pearl et al. 1990) and it is therefore possible that Uranus' thermal gradient is non-adiabatic and significant regions of its interior are non-convective. The magnetic fields of Uranus and Neptune provide additional information about their interiors (Ness et al. 1986, 1989; Stanley and Bloxham 2004, 2006). The magnetic fields require highly electrically conducting fluid regions, probably dominated by water and situated at relatively shallow depths, to account for the multipolar nature of the fields. \par

Two main methods have been used in modeling Uranus and Neptune. The first assumes that the planets consist of three layers: a core made of 'rocks' (silicates, iron), an 'icy' shell  (H$_2$O, CH$_4$, NH$_3$, and H$_2$S), and a gaseous envelope (composed of H$_2$ and He with some heavier components). This approach uses physical equations of state (EOSs) of the assumed materials to derive a density (and associated pressure and temperature) profile that best fits the measured gravitational coefficients. The physical parameters of the planets, such as mass and equatorial radius, are used as additional constraints. The masses and compositions of the three layers are modified until the model fits the measured gravitational coefficients. The models typically assume an adiabatic structure, with the adiabat being set to the measured temperature at the 1 bar pressure-level (Hubbard et al., 1991; Podolak et al., 1995). Although this method has succeeded in finding a model that fits both J$_2$ and J$_4$ for Neptune, no model of this type has been found that fits both J$_2$ and J$_4$ of Uranus. For example, Podolak et al. (1995) found that in order 
to fit the observed parameters for Uranus, it was necessary to assume that the density in the ice shell was 10\% lower than given by then-current equations of state. In addition, the ratio of ice to rock in this model was 30 by mass, roughly 10 times the solar ratio. Podolak et al. (1995) point out that Uranus' lower density might be explained by higher internal temperatures if the planet is not fully adiabatic. A non-adiabatic structure for Uranus is an appealing option since it suggests an explanation for the low heat flux of the planet in terms of an interior that is not fully convective.  \par

A second approach to model the interiors of Uranus and Neptune makes no a priori assumptions regarding planetary structure and composition. The radial density profiles of Uranus and Neptune that fit their measured gravitational fields are derived using Monte Carlo searches (Marley et al. 1995, Podolak et al. 2000). This approach is free of preconceived notions about planetary structure and composition and is not limited by the equations of state of assumed materials. Once the density profiles that fit the gravitational coefficients are found, conclusions regarding their possible compositions can be inferred using theoretical EOSs (Marley et al. 1995). \par 

In this paper we apply the method previously used in our models of Saturn (Anderson and Schubert 2007; Helled et al. 2009a) to derive continuous radial density and pressure profiles that fit the 
mass, radius, and gravitational moments of Uranus and Neptune. The use of a smooth function for the density with no discontinuities allows us to test whether Uranus and Neptune could have interiors with no density (and composition) discontinuities. Section 2 summarizes the models and results.  In section 3 we use physical  
equation of state tables to infer what these density distributions 
imply about the internal composition of Uranus and Neptune. Conclusions are discussed in section 4.  \par

\section{Interior Model: Finding Radial Profiles of Density and Pressure}

The procedure used to derive the interior model is described in detail in Anderson and Schubert (2007) and Helled et al. (2009a). 
The method is briefly summarized below. \par   
The gravitational field of a rotating planet is given by
\begin{eqnarray}
U &=& \frac{G M}{r} \left( 1 - \sum_{n=1}^\infty \left( \frac{a}{r} \right)^{2 n} \mathrm{J}_{2 n} \mathrm{P}_{2 n} \left( \cos \theta  \right)  \right) + \frac{1}{2} \omega^2 r^2 \sin^2 \theta .
\label{U}
\end{eqnarray}
where (r, $\theta, \phi$) are spherical polar coordinates, $G$ is the gravitational constant, $M$ is the total planetary mass, and $\omega$ is the angular velocity of rotation. 
We assume that the planets rotate as solid bodies with Voyager rotation periods (Table 1), although this assumption is a simplification since the interior rotation profiles of Uranus and Neptune are actually poorly known and could be more complex  (Helled et al. 2010). The potential $U$ is represented as an expansion in even Legendre polynomials P$_{2 n}$ (Kaula 1968; Zharkov and Trubitsyn 1978).  The planet is defined by its total mass, equatorial radius $a$ at the 1 bar pressure level, and harmonic coefficients J$_{2 n}$, which are inferred from Doppler tracking data of a spacecraft in the planet's vicinity. \par   

The measured gravitational coefficients of Uranus and Neptune are listed in Table 1. The observed gravitational coefficients J$_2$, J$_4$ correspond to the arbitrary reference equatorial radii R$_{\text{ref}}$ of 26,200 km and 25,225 km for Uranus and Neptune, respectively (Jacobson et al. 2006; Table 1). 
Another physical property that is used in the interior model is the equatorial radius. 
For Uranus the radio occultation of Voyager 2 yielded two radii on ingress and egress. These were nearly equatorial occultations and they provided essentially direct measurements of the planet's equatorial radius. Uranus' equatorial radius was found to be 25,559 $\pm$ 4 km. 
Neptune's occultation geometry was not equatorial. The geometry of Voyager 2 radio-occultation measurements was such that egress data were more difficult to interpret, resulting in one reliable planetocentric radius measurement on egress at a latitude of 42.26$^o$ S (Tyler et al., 1989; Lindal 1992, Figure 7). The radius at this latitude was found to be 24,601 $\pm$ 4 km. 
Lindal (1992) derived an equatorial radius of 24,766 $\pm$ 15 km for Neptune's 1 bar isosurface using wind velocities (Smith et al., 1989), with the large error reflecting the uncertainties in the extrapolation of the occultation measurement  to the equator and to the pole. 
A combination of stellar occultation measurements at the microbar pressure level (e.g., Hubbard et al. 1987) with a model atmosphere such as the one presented by Lindal (1992, Figures 3 and 4) can be used to derive the planetary shape at the 1 bar pressure level. For Neptune, such an extrapolation results in a very good agreement with the equatorial radius reported by Lindal. However, the extrapolation from 1 $\mu$bar to 1 bar assumed a quiescent hydrostatic atmosphere and did not take into account the wind systems between these atmospheric pressure levels. Following Helled et al. (2010) we adopt an equatorial radius of 24,764 $\pm$ 15 km for Neptune, an equatorial radius which is similar to the one reported by Tyler et al. (1989), and only 2 km smaller than the one derived by Lindal (1992). The planetary shape (mean radius) derived using Lindal's radii is different from the one found in our models. The difference is caused by the fact that we derive the planetary shape with the centrifugal potential defined solely by Voyager's solid-body rotation periods for both Uranus and Neptune, while Lindal derived the planetary shapes of the 1 bar pressure level including the distortion caused by the winds for both planets (see details in Helled et al., 2009b, 2010). \par

{\bf [Table. 1]}\\

We calculate the geoid radius to fifth order in the smallness parameter m (defined below, Table 2) and extrapolate the harmonics J$_2$ and J$_4$ to even harmonics J$_6$, J$_8$ and J$_{10}$ by a linear function in log(J$_{2i}$) versus log(2i) to insure that the precision of the geoid calculation is better than $\pm$ 1.0 km (Helled et al. 2010).  
The mean density $\rho_0$ of the planet is defined as the total mass divided by the volume of the fifth order reference geoid. The smallness parameter $m$ follows from its definition $m=\omega^2R^3/GM$, the planet's mean radius $R$ is the radius of an equivalent sphere with the mean density of the planet (Zharkov and Trubitsyn, 1978). The similar smallness parameter $q$, given by $\omega^2 a^3/GM$, is used for the calculation of the reference geoid. 
The characteristic pressure in the interior is defined by $p_0$=GM$\rho_0/R$ (Zharkov and Trubitsyn 1978). Given the parameters of Tables 1 and 2, the normalized mean radius $\beta$ is defined by $s/R$, where $s$ is the mean radius of an interior level surface, the normalized mean density $\eta(\beta)$ is $\rho(s)/\rho_0$, and the normalized pressure $\xi(\beta)$ is $p(s)/p_0$. The mean radii of Uranus and Neptune are found to be 25388.2 km and 24659.0 km, respectively. \par

For a given value of the smallness parameter $m$ and a density distribution $\eta(\beta)$, the level surfaces for constant internal potential can be evaluated and the surface harmonics can be computed from a series approximation in $m$ to the equation (Zharkov and Trubitsyn, 1978),
\begin{equation}
Ma^n\mathrm{J}_n = - \int_\tau \rho(r) r^n \mathrm{P}_n(\cos \theta) d \tau,
\label{Jn}
\end{equation}
where the integration is carried out over the volume $\tau$. 
We represent the internal density distribution by a single 6$^{th}$ degree polynomial with the first degree term missing  so that the derivative of the density goes to zero at the center.   The data to be satisfied by the interior model consist of the gravitational harmonics J$_2$, J$_4$, and the measured mass and mean radius of the planet. We start with a guess for $\eta(\beta)$, compute the level surfaces in the interior, and then evaluate the harmonic coefficients $J_2$, $J_4$, and $J_6$ at the surface for $\beta$ equal to unity. The differences between the calculated coefficients and the measured values are used to correct the density function, and the process is iterated to convergence.  For better matching of the interior polynomial to the atmosphere, we use the model atmospheres in Tables 10.2 and 11.2 of Lodders and Fegley (1998), and derive least-squares normal equations for the polynomial coefficients (see Helled et al. 2009a for details). Once a density distribution that matches the observed gravitational coefficients is found, the pressure in the interior is obtained by integration of the equations of hydrostatic equilibrium and mass continuity. \par
 
We use the level surface theory to third order and evaluate the gravitational harmonics $J_2$, $J_4$, and $J_6$ for an assumed density distribution $\eta(\beta)$. We iterate the polynomial coefficients until a best fit to the gravitational harmonics and the atmospheric data is obtained. The fit to the measured gravitational harmonics is not perfect because there is a trade-off between the fit to the atmosphere and the fit to the harmonics. The model results are listed in Table 2.  
The calculated gravitational coefficients $J_{2n}$ are normalized to the calculated mean radii $R$. For comparison, the observed gravitational coefficients normalized to the mean radii $\bar{J}_{2n}$ are listed as well.  
The fact that the two data sets are satisfied well within their respective standard errors lends credibility to the interior density distribution.\par

{\bf [Table. 2]}\\

{\bf [Figures 2, 3]}\\

Figure 2 presents the radial density profiles of Uranus (gray curve) and Neptune (black curve) found from the interior models. 
Neptune's density is found to be higher than Uranus' in agreement with previous models. It has been suggested that the higher density results from Neptune's greater compression (higher mass) and not from compositional differences (Hubbard et al. 1991; Marely et al. 1995, Podolak et al. 1995). Figure 3 shows the radial pressure profiles of Uranus and Neptune. The gray and black lines correspond to Uranus and Neptune, respectively. Figure 3 combines the results in the previous figures to produce pressure-density relations for Uranus and Neptune. The pressure-density profiles can be referred to as empirical EOSs for the planets; the EOSs of the two planets are similar. 

{\bf [Figure 3]}\\
Below we use the derived pressure-density profiles to investigate the possible compositional structures of the planets using physical equations of state for hydrogen, helium, rock (SiO$_2$), and ice (H$_2$O).

\section{Composition and Structure}

In this section we use the empirical $\rho$-p relations for compositional interpretation. 
Although Uranus and Neptune are known as 'icy planets', there is no direct evidence that they actually consist of a significant amount of ices. 
We consider two different possible compositions. The first is a mixture of rock (represented by SiO$_2$) with a hydrogen and helium mix (in the solar ratio), and the second a mix of ice (H$_2$O) with solar hydrogen and helium. In reality, the rocks and ices are likely to coexist, however, considering both materials will introduce still another free parameter in the model. For simplicity, and in order to minimize the number of free parameters, we separate the two cases. For hydrogen and helium we use SCVH EOS (Saumon et al., 1995). This EOS is based on free energy minimization, and is currently the one most commonly used for astrophysical applications. \par 

To describe the heavy elements in the interior model we use the EOS of silicon dioxide (SiO$_2$) and of water (H$_2$O) that are based on tables kindly supplied by D. Young (Young and Corey 1995; More et al.,~1988). 
The density of the mixture is calculated by the ``additive-volume rule'' (Saumon at
al.,~1995),
\begin{equation}
\frac {1}{\rho(P,T)} = \frac{X}{\rho_H} + \frac{Y}{\rho_{He}} + \frac{Z}{\rho_Z}, 
\end{equation}
where $X$ is the hydrogen mass fraction, $Y$ is the mass fraction of helium defined by $Y\equiv 1-X-Z$, $Z$ is the mass fraction of the high-Z material, and $\rho_i$ is the density of each component.\\
The entropy of the mixture is given by,
\begin{equation}\label{ent}
S(P,T)= XS_H(P,T)+YS_{He}(P,T)+ZS_{Z}(P,T)+S_{mix}(P,T),
\end{equation}
where $S_H$, $S_{He}$ and $S_{Z}$ are the entropy of hydrogen, helium, and silicon dioxide, respectively. $S_{mix}$ is the entropy of the mixture given by $S_{mix} = {k_B}(NlnN - \sum N_ilnN_i)$, where $N$ is the total number of particles (including free electrons), $N_i$ is the total number of particles of component $i$ (number of nuclei and electrons), and $k_B$ is Boltzmann's constant.  
When computing a 
mixture in our model, we first use the EOS of each individual
material and then combine it with the others by using the
additive-volume rule to compute the total density and the energy
by using the entropy as given by eq. 4. \par

\subsection{Uranus}
At a pressure of 1 bar our Uranus model has a density of 3.5$\times10^{-4}$\,g cm$^{-3}$. An ideal gas under these conditions and at a temperature of T = 75 K will have a mean molecular weight of 2.2, which is very close to the value for a solar mix of hydrogen and helium. We take this as our starting point. The temperature must increase toward the center of the planet. For an adiabat through a Debye solid, the temperature and density are related by 
\begin{equation}\label{adiab}
T=C\rho^\gamma
\end{equation}
where $C$ and $\gamma$ are constants that depend (among other things) on composition. Though we have noted above that the temperature gradient in Uranus need not be adiabatic we nevertheless use eq.\,\ref{adiab} as a simple and a representative formula for the internal thermal state. \par

The central pressure in the Uranus model is 5.93 Mbar, so that 
unless the temperature at the center is significantly higher than 10$^4$ K, the density will be 
very close to 7 g cm$^{-3}$. The model density is 4.42 g cm$^{-3}$, so the center of the planet 
cannot be pure SiO$_2$, but must contain an admixture of lighter material. The central density 
and pressure can be fit, at T = 10$^4$ K with a mixture of H, He and SiO$_2$ where H and He are 
in solar proportions (X = 0.112, Y = 0.038) and SiO$_2$ has a mass fraction Z = 0.85. A more careful calculation using eq. 5 to estimate the temperature also gives Z = 0.85.  
Of course it is more likely that the composition of both cores is actually a mixture 
of rock and ice, but, as Podolak et al.  (1991, Table 1) have shown, a mixture of H, He, 
and rock, will mimic the pressure-density behavior of ice. \par

Since the pressure-density relation near the surface is well-matched by a solar-composition mixture of H and He, we consider a compositional model where the mass fraction Z of SiO$_2$ increases from zero near the surface to its value at the center. Lacking any guidance as to the functional form of this increase, we assume a linear increase in Z with log $\rho$ from Z = 0 at log $\rho= -0.455$ to Z $= 0.85$ at log $\rho$ = 3.646. To estimate the temperature we use eq. 5 with $\gamma$= 0.5, which is the asymptotic value of $\gamma$ for high pressure. At lower pressures $\gamma$ will be higher. As a result, our procedure underestimates the temperature in the lower pressure regions of the planet, but it gives us a feeling of whether the proposed compositional model makes any sense.  We call this model case I. \par

We repeat the above calculations with H$_2$O representing the high-Z material. Figure 5 shows the comparison between a 6$^{th}$ order polynomial that fits all the gravitational data (black solid curve) and the compositional model described above with rock (black dashed curve) and ice (gray dashed curve). As can be seen from the figure, the behavior of the calculated EOS curves are very close to each other although the curve representing a mixture of hydrogen and helium with ices is slightly closer to curve of the empirical $\rho$-p relation.  Considering the crudeness of the approximations involved, the fits are quite good. It is not surprising that the density at intermediate pressures is too high. This is most likely due to our underestimate of the temperature in that region. Certainly the details of the temperature gradient and the compositional gradient can be adjusted to provide a better fit. In any case, it seems quite clear that a Uranus model with an adiabat-like temperature gradient and a continuous increase in Z toward the center can be made to fit the observed parameters of the planet.\par 

We next investigate the possibility of a different compositional configuration. We consider a three-layer model, where the outermost layer has Z constant and equal to its value at the surface; the innermost layer has Z constant and equal to its value at the center, and the middle layer has Z varying linearly with $\log \rho$ between the surface value and the central value. The beginning and end of this central "transition region" is determined by getting a good fit to the empirically determined EOS.  We call this model case II.  This model gives a surprisingly good fit to the empirical curve and is presented as black  (rock) and gray (ice) dots in Figure 5.  Remarkably the transition region begins at a temperature of around 1500\,K, which is just the region where silicates begin to vaporize.  The overall composition of Uranus is essentially the same for case I and case II (see table 3).

\subsection{Neptune}
The best fit polynomial gives a 1-bar density for Neptune of 4.38$\times 10^{-4}$\,g cm$^{-3}$. For an ideal gas at this pressure and at T = 75 K, the mean molecular weight is 2.7. This is significantly higher than the 2.3 appropriate for a solar mix of H and He. Although the vapor pressure of SiO$_2$ at this temperature is extremely low, for the purposes of this comparison we assume that SiO$_2$ vapor supplies the additional molecular weight. Its mass fraction at the 1-bar level must then be Z = 0.0073.  Again we assume that the temperature is given by eq. 5, so that the central temperature is 8.1$\times 10^3$ K. At this temperature, and at a central pressure of 8.22 Mbar as determined from the polynomial fit, the central density of 5.15 g cm$^{-3}$ is fit by Z = 0.82. This is very similar 
to the value determined for Uranus. \par

If we assume that Z increases linearly in log $\rho$ as before (case I), we find that the pressure-density relation is given by the dashed curve in fig. 6. This is quite close to the pressure-density relation determined by a polynomial fit to Neptune's observed parameters (black solid curve in fig. 6). A case II model of Neptune gives similar results.  Here the transition region begins at a somewhat lower temperature, $\sim 1400$\,K.  \par

For both Uranus and Neptune the polynomial fit and the proposed compositional model differ in the same way. For case I both Uranus and Neptune have too high a density at  intermediate pressures, indicating that the assumed temperature there is too low or Z is too high. Nearer to the center the density is somewhat too low for both planets. In this region the temperature dependence of the pressure is small, so it is more likely that a somewhat faster than linear increase in Z is indicated. In any case, essentially the same compositional model works for both Uranus and Neptune. For case II both planets show the same overall structure.  The transition region for both planets begins at around 1500\,K.  The big difference between the two planets is that Neptune requires a non-solar envelope while Uranus is best matched with a solar composition envelope.  A similar result was found by Podolak et al. (1995), but it was not clear how much of this difference was due to the detailed assumptions of the internal structure and temperature distribution.  We find a qualitatively similar result simply from a polynomial fit to the observed gravity field.  Finally, we note that both H$_2$O and SiO$_2$ match the polynomial fit equally well. \par

The model fits can be used to derive the relative amounts of each component (mass fractions). For case I we find that when SiO$_2$ is used, Uranus's interior is found to consists of 18.1\% hydrogen, 6.2\% helium, and 75.7\% rock. Neptune's composition is found to be 18.1.\% hydrogen, 6.2\% helium, and 75.8\% rock. When H$_2$O is used, Uranus' interior is found to consist of 8.48\% and 2.88\% of hydrogen and helium, respectively, and 88.6\% of ice.  Neptune's composition is found to be 8.0\% hydrogen, 2.7\% helium, and 89.3\% ices.
For case II we find that Uranus consists of 16.4\% hydrogen, 5.56\% helium, and 78.1\% rock. When H$_2$O is used, Uranus' interior is found to consist of 6.41\% and 2.18\% of hydrogen and helium, respectively, and 91.4\% of ice.  Neptune's composition is found to be 17.5.\% hydrogen, 6.94\% helium, and 76.6\% rock. When water is used to represent the high-Z material, Neptune's interior is found to consist of 7.19\% and 2.44\% of hydrogen and helium, respectively, and 90.4\% of ice. 
The results are summarized in table 3. It is clear that the compositions of the planets are similar, and as expected, the high-Z mass fraction increases when a lighter material (H$_2$O) is used to represent the heavy elements. 
 
{\bf Table 3}\\ 
 
The compositions listed above are not meant to represent the actual compositions of Uranus and Neptune but are only the compositions derived under our simple interpretation of the polynomial fit.  In addition, as we mentioned earlier, the planets are likely to consist of a mixture of both ice and rock. Clearly more detailed theoretical models are necessary.  \par

\section{Discussion and Conclusions}
We present phenomenological pressure-density profiles for Uranus and Neptune. These models show that the gravitational fields (J$_2,$ J$_4$) of both Uranus and Neptune could be reproduced with continuous radial density distributions (no density jumps). In addition, the internal structures of the planets are found to be similar.  We use equations of state for hydrogen and helium, and high-Z material (SiO$_2$ or H$_2$O) to interpret the empirical EOS in terms of compositional structure. 
Our analysis suggests that Uranus and Neptune could have similar composition and structure. We find that the empirical EOS for both planets can be fit by assuming a gradual increase of the heavier material toward the center, with hydrogen and helium being kept at solar ratio. For case I, when SiO$_2$ (rock) is used to represent the high-Z material in the interior, the masses of heavy elements are found to be 10.9 and 12.9\,M$_{\oplus}$ for Uranus and Neptune, respectively. 
The innermost regions of both Uranus and Neptune cannot be fit to the empirical EOS by pure rock, but by $\sim$ 82\% rock by mass for both planets with the rest being a mixture of hydrogen and helium in solar ratio. \par

When H$_2$O is used to represent high-Z material the masses of heavy elements are found to be higher due to the lower density of water compared with rock. The heavy element masses are found to be $\sim$ 12.8 and 15.2\,M$_{\oplus}$ for Uranus and Neptune, respectively. Even when water is used, the planetary centers are found to contain gases in addition to the high-Z material, though the Z mass fraction at the centers are found to be above 90\% for both planets. Again, the results imply that the interiors of Uranus and Neptune could have concentrations of high-Z material increasing gradually toward their centers. A similar situation might occur in the gas giant planets, Jupiter and Saturn, and we suggest that a gradual increase of the heavier elements should be considered when modeling giant planet interiors.  For case II the SiO$_2$ contents for Uranus and Neptune are 11.3 and 13.1\,M$_{\oplus}$, respectively, while for H$_2$O the values are 13.2 and 15.4\,M$_{\oplus}$.  \par 

The interior structures of Uranus and Neptune are poorly understood, and their interiors may be quite different from the 'traditional 3-layer' models. A  3-layer model is often used to model interior structures of Neptune/Uranus-like extrasolar planets, while the true interiors of Uranus and Neptune may differ significantly from such a structure. We therefore suggest that more flexible interior structures should be considered when modeling extrasolar planet interiors. 
 
\subsection*{Acknowledgments} 
R. H. and J. D. A acknowledge support from NASA through the Southwest Research Institute. M. P. acknowledges support from ISF 388/07. 
G. S. acknowledges support from the NASA PGG and PA programs.

\section{References}

Anderson, J. D. and Schubert, G. "Saturns Gravitational Field, Internal Rotation, and Interior Structure." Science 317: (2007) 1384---1387. \\
Guillot, T., 1999. A comparison of the interiors of Jupiter and Saturn. Planetary and Space Science, 47, p. 1183--1200. \\ 
Guillot, T. and  Gautier, D., 2007, Giant Planets, Treatise of Geophysics, vol. 10, Planets and Moons, Schubert G., Spohn T. (Ed.) (2007) 439-464. eprint arXiv:0912.2019. \\
Helled, R., G. Schubert, and J. D. Anderson. ÒEmpirical models of pressure and density in Saturn's interior: Implications for the helium concentration, its depth dependence, and Saturn's precession rateÓ, Icarus, 2009(a), 199, 368--377. \\
Helled, R., Schubert, G. and Anderson, J. D. ÒJupiter and Saturn Rotation Periods.Ó Planetary and Space Science, 2009(b), 57, 1467--1473. \\
Helled, R., J. D. Anderson, and G. Schubert. ÒUranus and Neptune: Shape and Rotation". Icarus, 2010, in press.\\
Hubbard, W. B., 1968. Thermal structure of Jupiter. Astrophys. J., 152:745--754. \\
Hubbard, W. B., Nicholson, P. D., Lellouch, E., Sicardy, B., Brahic, A., Vilas, F., Bouchet, P., McLaren, R. A., Millis, R. L., Wasserman, L. H., Elais, J. H., Matthews, K., McGill, J. D., Perrier, C., 1987. "Oblateness, radius, and mean stratospheric temperature of Neptune from the 1985 August 20 occultation", Icarus, 72, 635--646. \\
Hubbard, W. B., Nellis, W. J., Mitchell, A. C., Holmes, N. C., McCandless, P. C., and
Limaye, S. S. (1991). Interior structure of Neptune - Comparison with Uranus. Science,
253, 648--651.\\
Lindal, G. F. ÒThe atmosphere of Neptune - an analysis of radio occultation data acquired with Voyager 2.Ó Astron. Jour. 103: (1992) 967Ð982. \\
Lindal, G. F., Sweetnam, D. N., and Eshleman, V. R. 1985, The atmosphere of Saturn - an analysis of the Voyager radio occultation measurements. AJ, 90, 1136--1146.\\
Lindal, G. F., J. R. Lyons, D. N. Sweetnam, V. R. Eshleman, and D. P. Hinson. ÒThe atmosphere of Uranus - Results of radio occultation measurements with Voyager 2.Ó Jour. Geophys. Res. 92: (1987) 14,987Ð15,001.\\
Lodders, Katharina, and Bruce Fegley, Jr. The Planetary Scientists Companion. New York Oxford: Oxford University Press, 1998. \\
Lyon, S. P. and Johnson, J. D., 1992. SESAME: The Los Alamos National Laboratory equation of state database. website, http://t1web.lanl.gov/doc/SESAME-3Ddatabase1992.html. \\
Marley, M. S., G—mez, P., and Podolak, M., 1995. Monte Carlo interior models for Uranus and Neptune. Journal of Geophysical Research, 100, 23349--23354.\\
More, R.M, Warren, D.A., Young, D.A \& Zimmerman, G.B. ,1988. A
new quotidian equation of state (QEOS) for hot dense matter.
Physics of Fluids, 31,10, 3059--3078.\\
Ness, N. F., Acuna, M. H., Behannon, K. W., Burlaga, L. F.,  Connerney, J. E. P., Lepping, R. P., 1986. Magnetic fields at Uranus. Science, 233, pp. 8--89.\\
Ness, N. F., Acuna, M. H., Burlaga, L. F., Connerney, J. E. P.; Lepping, R. P., 1989. Magnetic fields at Neptune. Science, 246, pp. 1473--1478. \\
Pearl, J. C., Conrath, B. J., Hanel, R. A., Pirraglia, J. A., and Coustenis, A., 1990. The albedo, effective temperature, and energy balance of uranus, as determined from voyager iris data. Icarus, 84:12--28. \\
Podolak, M. and Hubbard, W. B., 1998. Ices in the giant planets. In . M. F. B. Schmitt, C. de Bergh, ed., Solar System Ices, pp. 735--748. \\
Podolak, M., Hubbard, W.B. and Stevenson, D.J. Models of Uranus' interior and magnetic field. Uranus, ed. J. Bergstrahl and M. Matthews, Un. Arizona Press, pp.29-61, 1991.\\
Podolak, M., Weizman, A., and Marley, M., 1995. Comparative models of Uranus and Neptune. Planetary and Space Science, 43, 1517--1522.\\
Podolak, M., Podolak, J. I., and Marley, M. S., 2000. Further investigations of random models of Uranus and Neptune. Planet. \& Sp. Sci., 48:143--151. \\
Smith, B. A., L. A. Soderblom, D. Banfield, C. Barnet, R. F. Beebe, A. T. Bazilevskii, K. Bollinger, J. M. Boyce, G. A. Briggs, and A. Brahic. ÒVoyager 2 at Neptune - Imaging science results.Ó Science 246: (1989) 1422Ð1449. \\
Smith, B. A., L. A. Soderblom, R. Beebe, D. Bliss, R. H. Brown, S. A. Collins, J. M. Boyce, G. A. Briggs, A. Brahic, J. N. Cuzzi, and D. Morrison. ÒVoyager 2 in the Uranian system - Imaging science results.Ó Science 233: (1986) 43Ð64. \\
Stacey, Frank D. Physics of the Earth. Brisbane: Brookfield Press, 1992. \\
Stanley, S. and Bloxham, J., 2004. Convective-region geometry as the cause of Uranus' and Neptune's unusual magnetic fields. Nature, 428, pp. 151--153.\\
Stanley, S. and Bloxham, J., 2006. Numerical dynamo models of Uranus' and Neptune's magnetic fields. Icarus, 84, p. 556--572.\\
Saumon, D., Chabrier, G., \& Van Horn, H.M. 1995, An equation of
state for low-mass stars and giant planets. The Astrophysical Journal,
99, 713--741.\\
Saumon,D. \& Guillot,T., 2004. Shock Compression of Deuterium and
the Interiors of Jupiter and Saturn. The Astrophysical Journal, 609, 1170--1180.\\
Smith, B. A., L. A. Soderblom, D. Banfield, C. Barnet, R. F. Beebe, A. T. Bazilevskii, K. Bollinger, J. M. Boyce, G. A. Briggs, and A. Brahic., 1989. ÒVoyager 2 at Neptune - Imaging science results.Ó Science 246, 1422--1449. \\
Tyler, G. L., Sweetnam, D. N., Anderson, J. D., Borutzki, S. E., Campbell, J. K., Kursinski, E. R., Levy, G. S., Lindal, G. F., Lyons, J. R., Wood, G. E., 1989. Voyager radio science observations of Neptune and Triton. Science, 246, 1466--1473.\\
Young, D.A. \& Corey, E.M., 1995. A new global equation of state
model for hot, dense matter. J.Appl.Phys., 78:3748--3755.\\
Zharkov, V. N. and Trubitsyn, V. P. , 1974. Determination of the 
equation of state of the molecular envelopes of Jupiter and Saturn from their gravitational 
moments. Icarus, 21, 152--156. \\
Zharkov, V. N., and V.P. Trubitsyn. Physics of Planetary Interiors. Tucson: Pachart, 1978.


\begin{table}[h!]
\begin{center}
\begin{tabular}{lcccl}
\multicolumn{3}{c}{} \\
\hline
\hline
Parameter & Uranus & Neptune\\
\hline
P (rotation period) & 17.24h & 16.11h\\
GM (km$^3$ s$^{-2}$) & 5,793,964 $\pm$ 6& 6,835,100. $\pm$ 10 \\
R$_{\text{ref}}$ (km) & 26,200 & 25,225 \\
J$_2$ ($\times$10$^{6}$) &3341.29 $\pm$ 0.72&3408.43 $\pm$ 4.50\\
J$_4$ ($\times$10$^{6}$) & -30.44 $\pm$ 1.02& -33.40 $\pm$ 2.90\\
a (km)  & 25,559 $\pm$ 4 & 24,764 $\pm$ 15  \\
$\bar{J}$$_2$ ($\times$10$^{6}$) &3510.99 $\pm$ 0.72& 3536.51 $\pm$ 4.50\\
$\bar{J}$$_4$ ($\times$10$^{6}$) & -33.61 $\pm$ 1.02& -35.95$\pm$ 2.90\\
$q$ & 0.0295349 & 0.0260784 \\
$\rho_0$ (kg m$^{-3}$) & 1266.46 &1630.53 \\
$p_0$ (Mbar) & 2.89025 & 4.51959 \\
\hline
\end{tabular}
	\caption{{\small Physical data, taken from JPL database:  http://ssd.jpl.nasa.gov, Jacobson (2003), Jacobson et al. (2006). R$_{\text{ref}}$ is an arbitrary reference equatorial radius associated with the reported values of the measured gravitational harmonics J$_2$ and J$_4$.  $a$ is the equatorial radius at the 1 bar pressure level. $\bar{J}$$_2$ and $\bar{J}$$_4$ are the values taken by the gravitational coefficients when referenced to $a$ instead of R$_{\text{ref}}$. $q$, $\rho_0$ and $p_0$ are the smallness parameter, the characteristic density, and the pressure, respectively, as defined in the text. 
}}
\end{center}
\end{table}

\begin{table}[h!]
\begin{center}

\vskip 8pt
\begin{tabular}{l c c}
\hline
\hline
Parameter  & Uranus & Neptune 
\\
\hline

$J_2$ ($\times$10$^{6}$) & 3341.31 & 3408.65
\\
$J_4$ ($\times$10$^{6}$)& -30.66 & -30.97
\\
$J_6$ ($\times$10$^{6}$)& 0.4437 & 0.4329
\\
\hline
$\bar{J_2}$ ($\times$10$^{6}$) & 3511.01 & 3536.74
\\
$\bar{J_4}$ ($\times$10$^{6}$)& -33.85 & -33.34
\\
$\bar{J_6}$ ($\times$10$^{6}$)&  0.5148 & 0.4836
\\
\hline
\end{tabular} 
\caption{\label{geoid} 
Interior model results. $J_{2n}$ are the gravitational moments corresponding to the reference radii, R$_{\text{ref}}$, while $\bar{J_{2n}}$ correspond to the gravitational coefficients normalized by $a$ the equatorial radius at 1 bar. The values should be compared with the measured values given in Table 1.
}
\end{center} 
\end{table}

\begin{table}[h!]
\begin{center}

\vskip 8pt
\begin{tabular}{l c c}
\hline
\hline
 & SiO$_2$ & H$_2$O 
\\
\hline
Case I: Uranus & X = 0.181; Y = 0.0616; Z = 0.757 \vline &X = 0.0848; Y = 0.0288; Z = 0.886
\\
Case I: Neptune & X = 0.181; Y = 0.0615; Z = 0.758 \vline &  X = 0.0795; Y = 0.027; Z = 0.893  
\\
\hline
Case II: Uranus & X = 0.164; Y = 0.0556; Z = 0.781 \vline &X = 0.0641; Y = 0.0218; Z = 0.914
\\
Case II: Neptune & X = 0.175; Y = 0.0694; Z = 0.766 \vline &  X = 0.0719; Y = 0.0244; Z = 0.904  
\\
\hline
\end{tabular} 
\caption{\label{composition} 
Derived planetary compositions (given in mass fractions) for Uranus and Neptune. The two different columns correspond to different materials representing the heavy elements (SiO$_2$ and H$_2$O). The hydrogen to helium ratio (X/Y) is set to the protosolar value. Case I corresponds to a model assuming a linear increase in Z from the surface to the planet's center. Case II corresponds to an interior structure consists of three-layers with a constant composition in the central region (core) and the atmosphere with a middle 'transition region' with Z increasing toward the center (see text for details). The sum of the mass fractions do not exactly total to one because of numerical roundoff.}
\end{center} 
\end{table}

\newpage

\begin{figure}
    \centering
    \includegraphics[width=4.5in]{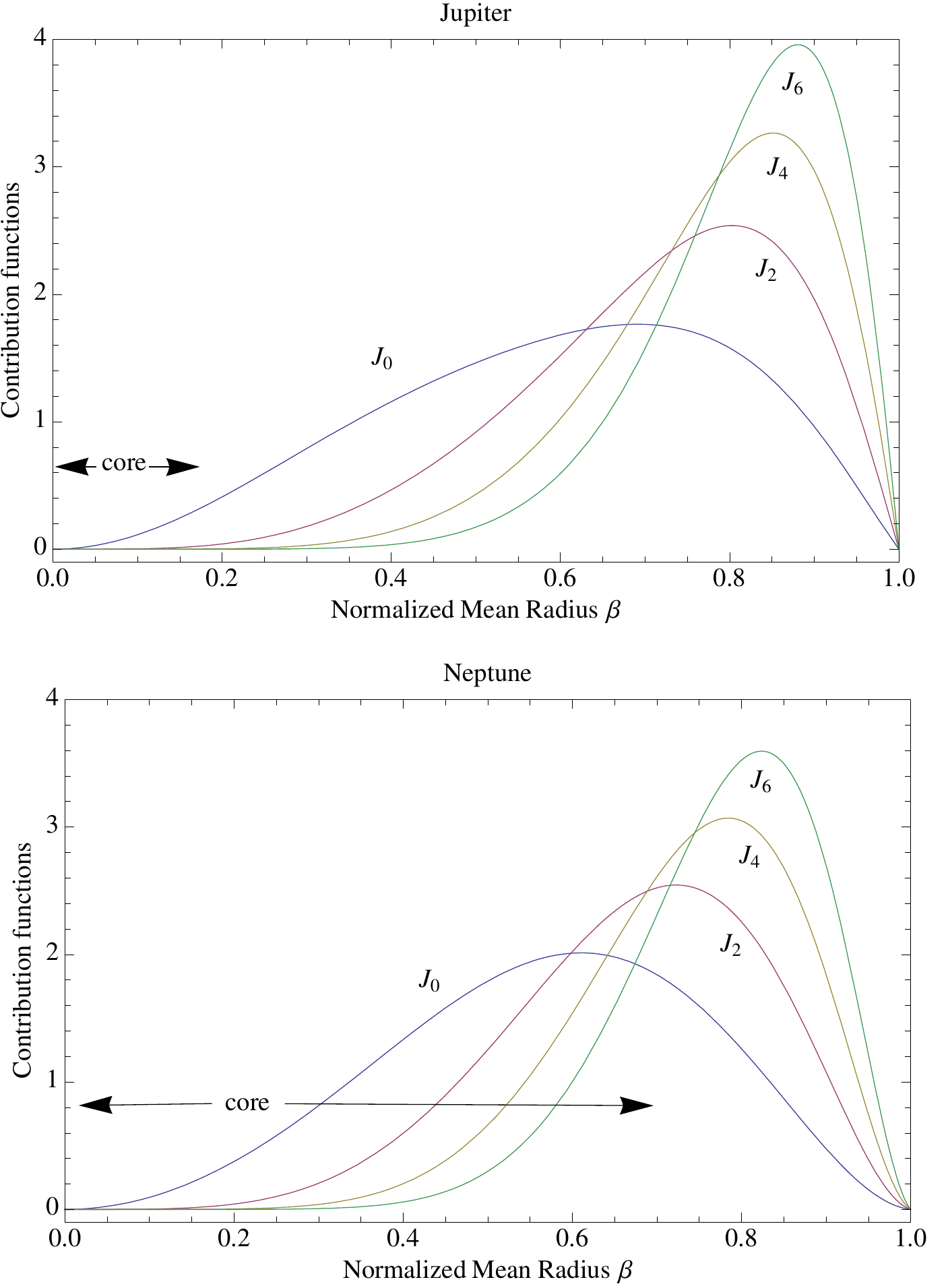}
    \caption[eos]{Normalized integrands of the gravitational moments (contribution functions) of Jupiter (top) and Neptune (bottom). The values are normalized to make the area under each curve equal unity. $J_0$ is equivalent to the planetary mass. The range of possible core sizes is indicated. Here, core designates a region of heavy elements below the H/He envelope. It is clear that Neptune's (Uranus') interior is better sampled by the gravitatioal harmonics compared to Jupiter (Saturn).}
\end{figure}

\begin{figure}
    \centering
    \includegraphics[width=5in]{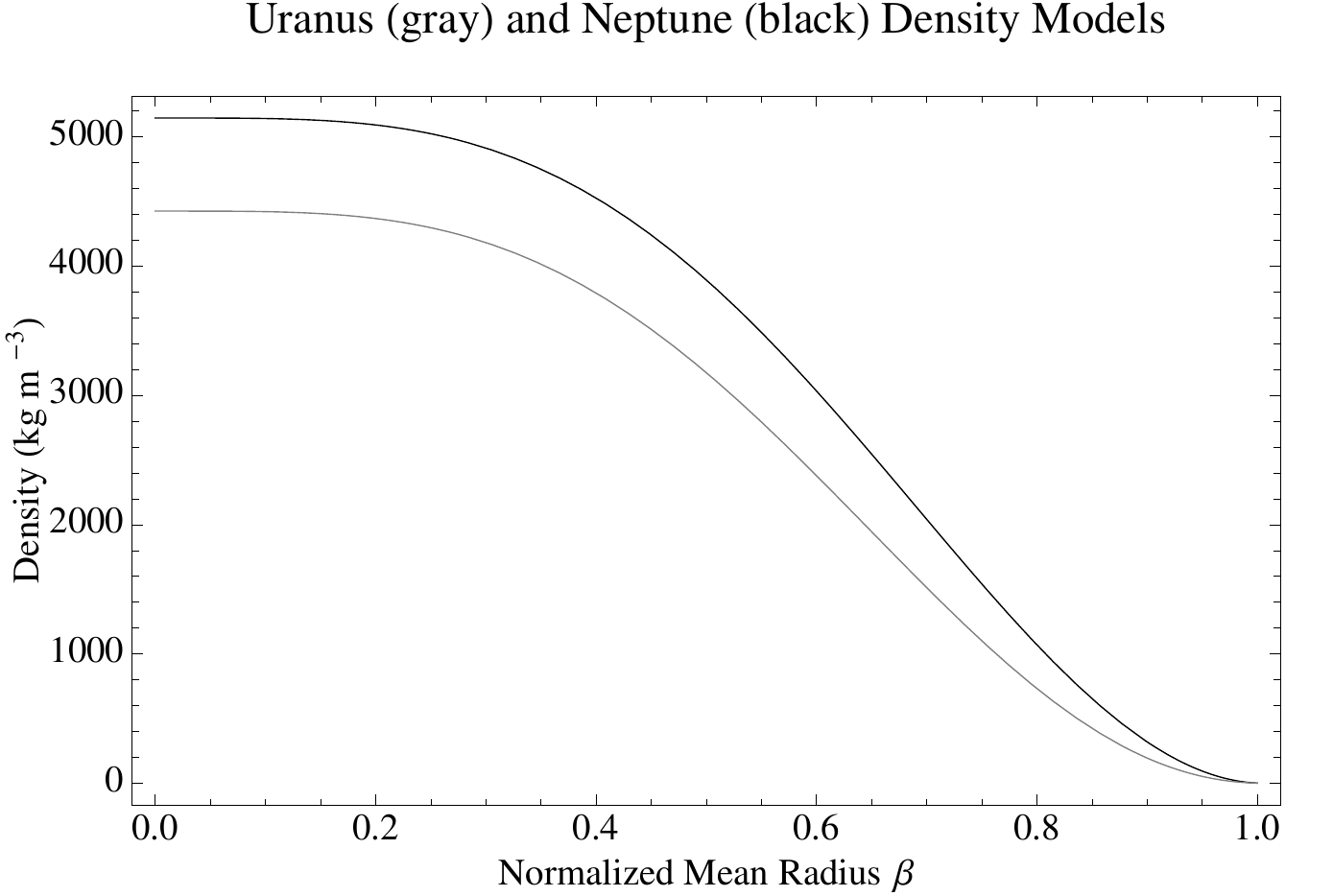}
    \caption[eos]{Uranus and Neptune density profiles as a function of normalized radius. }
\end{figure}

\begin{figure}
    \centering
    \includegraphics[width=5in]{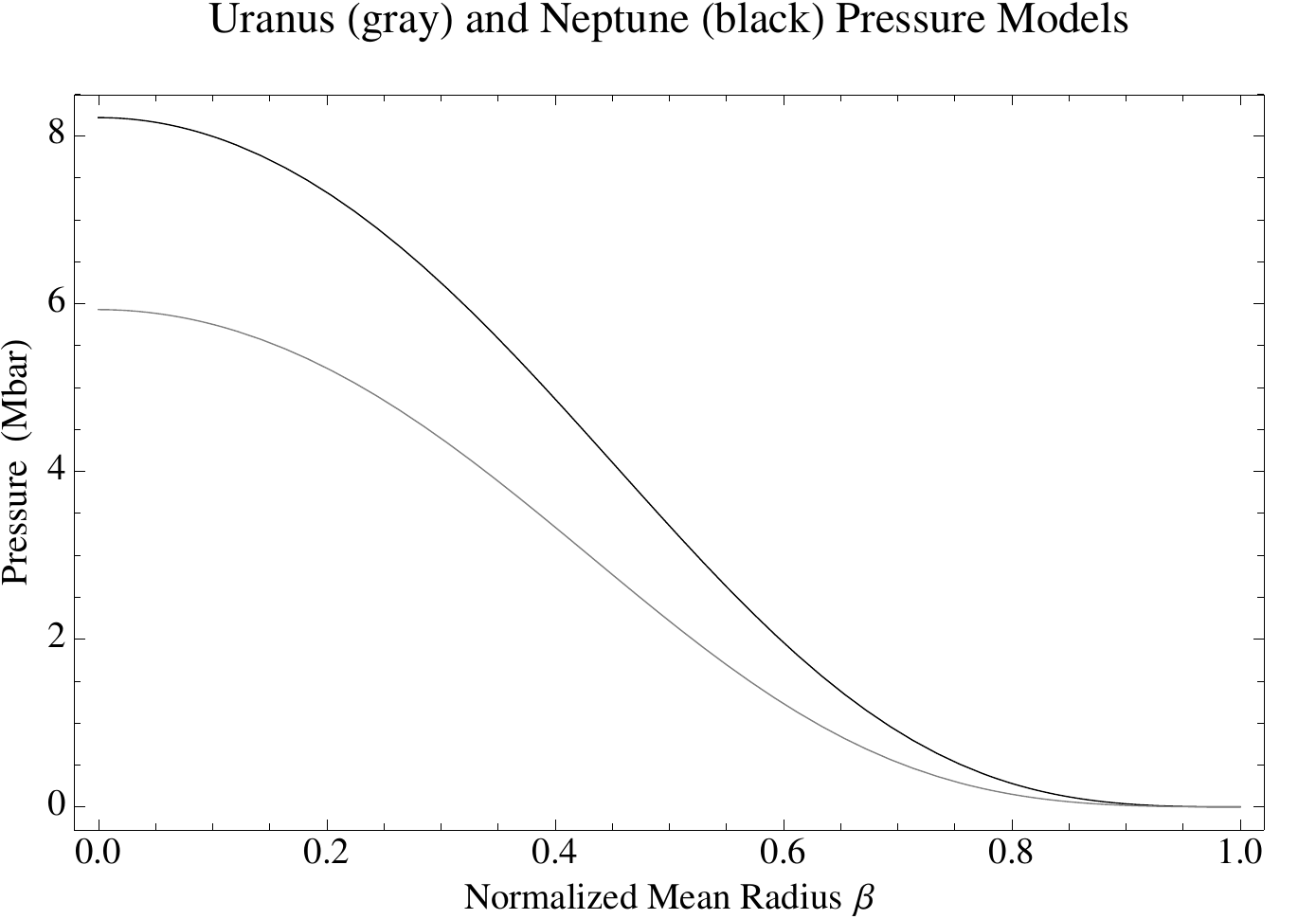}
    \caption[eos]{Uranus and Neptune pressure profiles as a function of normalized radius.}
\end{figure}

\begin{figure}
    \centering
    \includegraphics[width=5in]{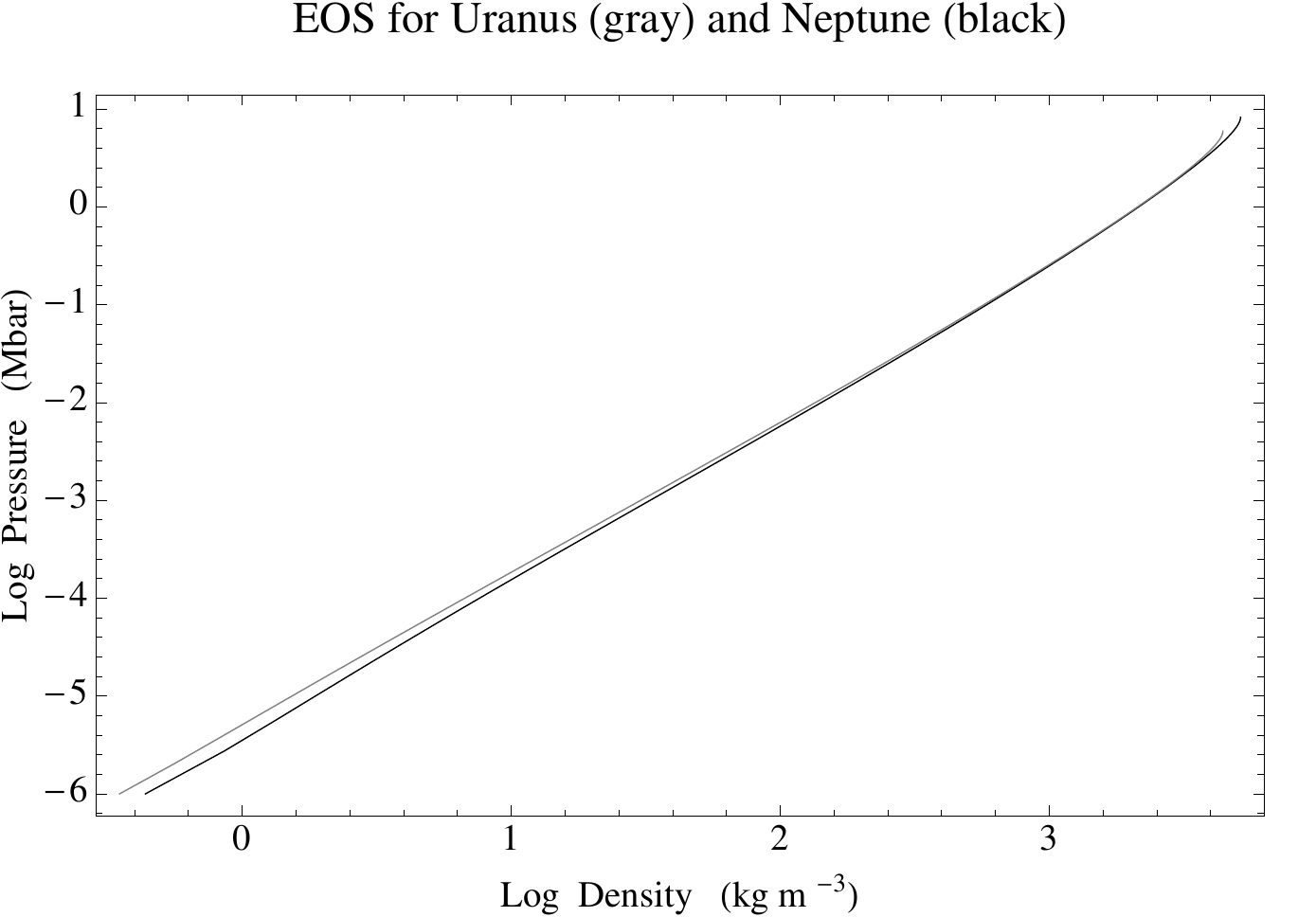}
    \caption[eos]{Uranus and Neptune empirical $\rho(p)$ relations. }
\end{figure}
\begin{figure}
    \centering
    \includegraphics[width=5.2in]{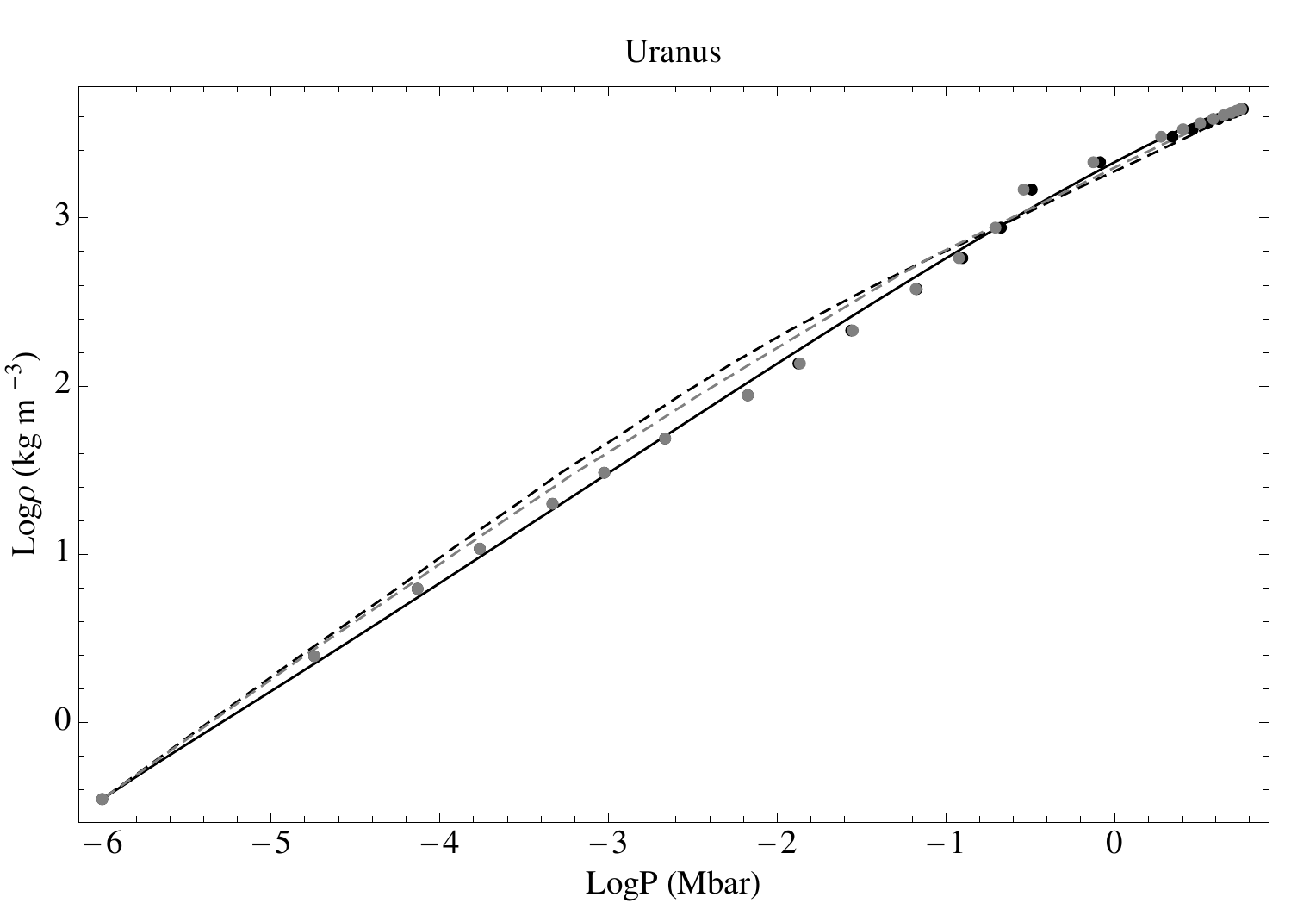}
    \caption[eos]{Pressure density relation for a Uranus model. The black solid curve is the polynomial 
that fits the gravitational data. The black and gray dashed curves are the compositional models described in 
the text taking the high-Z material to be SiO$_2$ and H$_2$O, respectively (Case I). The black and gray points are for Case II and correspond to rock and ice, respectively.}
\end{figure}
\begin{figure}
    \centering
    \includegraphics[width=5.2in]{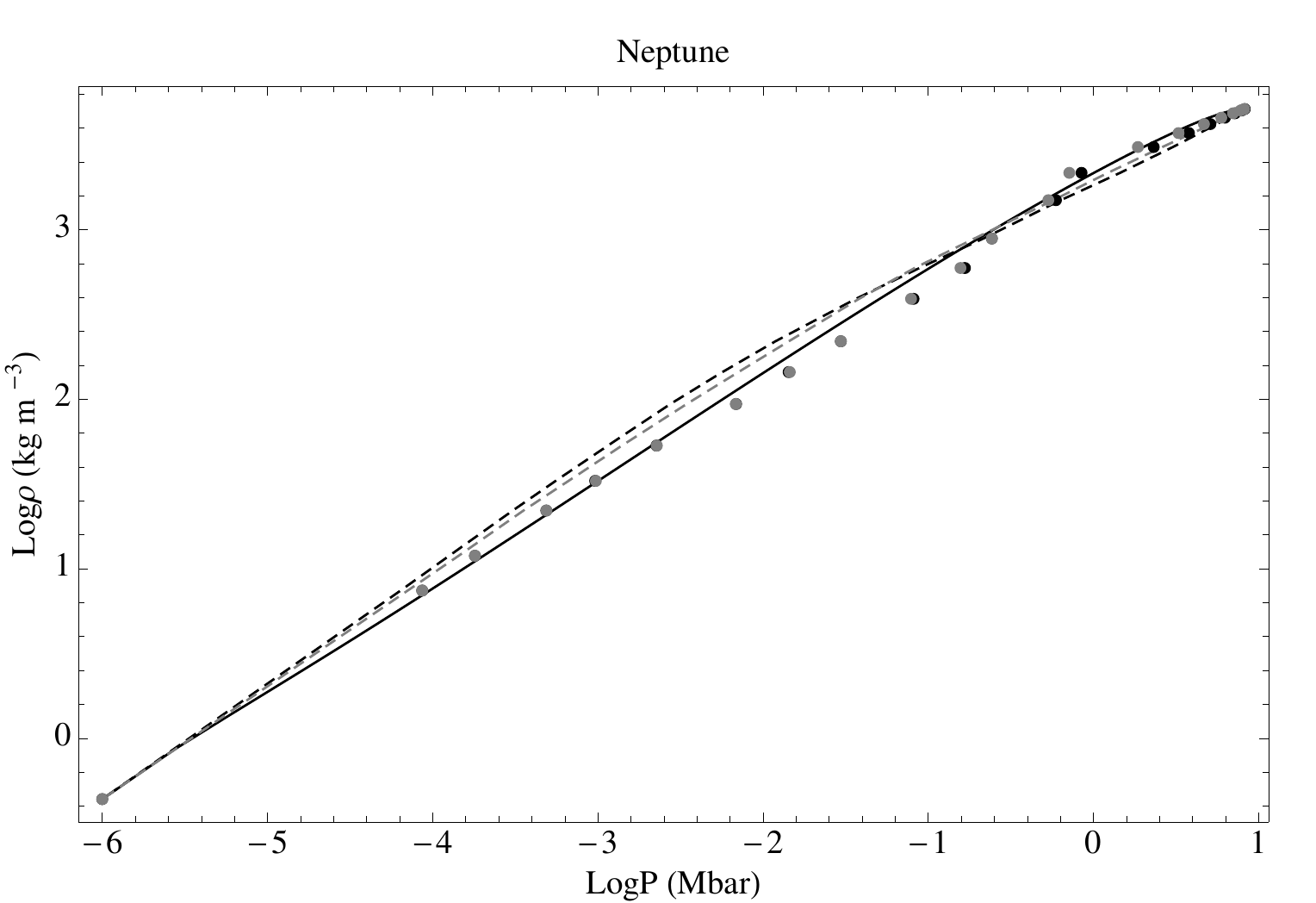}
    \caption[eos]{Pressure density relation for a Neptune model. The black solid curve is the polynomial 
that fits the gravitational data. The black and gray dashed curves are the compositional models described in 
the text taking the high-Z material to be SiO$_2$ and H$_2$O, respectively (Case I). The black and gray points are for Case II and correspond to rock and ice, respectively.}
\end{figure}

\end{document}